\documentclass{mpe_report}

\usepackage{psfig,graphicx,epsfig}
\usepackage{color}
\usepackage{amsmath,amssymb,epic,eepic,array}

\begin{document}

\pagenumbering{arabic}
\setcounter{page}{32}

\renewcommand{\FirstPageOfPaper }{ 32}\renewcommand{\LastPageOfPaper }{ 35}

\def\beq{\begin{equation}}
\def\enq{\end{equation}}


\title{511 keV line from millisecond pulsars in the Galactic center}
\author{Wei Wang}
\institute{Max--Planck--Institut f\"ur extraterrestrische Physik,
 Giessenbachstra{\ss}e, 85748 Garching, Germany \\
 e-mail: wwang@mpe.mpg.de}

\maketitle

\begin{abstract}
Observations of a strong and extended positron-electron
annihilation line emission in the Galactic center (GC) region by
the SPI/INTEGRAL are challenging to the existing models of
positron sources in the Galaxy. In this paper, we study the
possibility that pulsar winds from a millisecond pulsar population
in the GC produce the 511 keV line. Our preliminary estimations
predict that the e$^\pm$ annihilation rate in the GC is around
$5\times 10^{42} s^{-1}$, which is consistent with the present
observational constraints. Therefore, the e$^\pm$ pairs from
pulsar winds can contribute significantly to the positron sources
in the Galactic center region. Furthermore, since the diffusion
length of positrons is short in the magnetic field, we predict
that the intensity distribution of the annihilation line should
follow the distribution of millisecond pulsars, which should then
correlate to the mass distribution in the GC.

\end{abstract}

\section{511 keV electron-positron annihilation emission in the Galaxy}

Since the first detection (Johnson \& Haymes 1973) and subsequent
identification (Leventhal et al. 1978) of the Galactic 511 keV
annihilation line, the origin of the galactic positrons has become
a lively topic of scientific debate. With the launch of the
INTEGRAL gamma-ray observatory in 2002, the SPI board on telescope
provides a strong constraints on the morphology and intensity of
the 511 keV line emission from the Galactic center (Kn\"odlseder
et al. 2003; Jean et al. 2003). The data analyses show that the
line emitting source is diffuse, and that the line flux within
5$^\circ$ of the GC amounts to $\sim (9.9\pm 4) \times 10^{-4}{\rm
photon\ cm^{-2}\ s^{-1}}$ (Kn\"odlseder et al. 2003),
corresponding to a luminosity of $\sim 10^{36}{\rm erg\ s^{-1}}$.
The high line luminosity requires that the positron injection rate
in the GC should be around $(3-6)\times 10^{42}\ e^+{\rm s^{-1}}$.
Recently, analyses of the observational data by SPI/INTEGRAL with
deep Galactic center region exposure show that the spatial
distribution of 511 keV line appears centered on the Galactic
center (bulge component), with no contribution from a disk
component (Teegarden et al. 2005; Kn\"odlseder et al. 2005;
Churazov et al. 2005), and the positron injection rate is up to
$10^{43}\ e^+{\rm s^{-1}}$ within $\sim 6^\circ$.

The SPI observations present a challenge to the present models of
the origin of the galactic positrons, e.g. supernovae. Recently,
Cass\'e et al. (2004) suggested that hypernovae (Type Ic
supernovae/gamma-ray bursts) in the Galactic center may be the
possible positron sources. Moreover, annihilations of light dark
matter particles into $e^\pm$ pairs (Boehm et al. 2004) have been
proposed to be the potential origin of the 511 keV line in the GC.
Cheng et al. (2006) also propose that the continuous capture of
stars by the supermassive black hole at Sgr A* could explain the
morphology and intensity of the e$^\pm$ annihilation line. Here,
we suggest that a population of millisecond pulsars (MSP) can
significantly contribute to the 511 keV line in the GC.

\section{Motivations of a millisecond pulsar population in the GC}

Millisecond pulsars are old pulsars which could have been members
of binary systems and been recycled to millisecond periods, having
formed from low mass X-ray binaries in which the neutron stars
accreted sufficient matter from either white dwarf, evolved main
sequence star or giant donor companions. The current population of
these rapidly rotating neutron stars may either be single (having
evaporated its companion) or have remained in a binary system. In
observations, generally millisecond pulsars have a period $< 20$
ms, with the dipole magnetic field $< 10^{10}$ G. Figure 1 shows
the distribution of observed MSPs in our Galaxy, and they
distribute in two populations: the Galactic field (1/3) and
globular clusters (2/3). In the Galactic bulge region, there are
four globular clusters, including the famous Terzon 5 in which 27
new millisecond pulsars were discovered (Ransom et al. 2005).

\begin{figure}
\centerline{\psfig{file=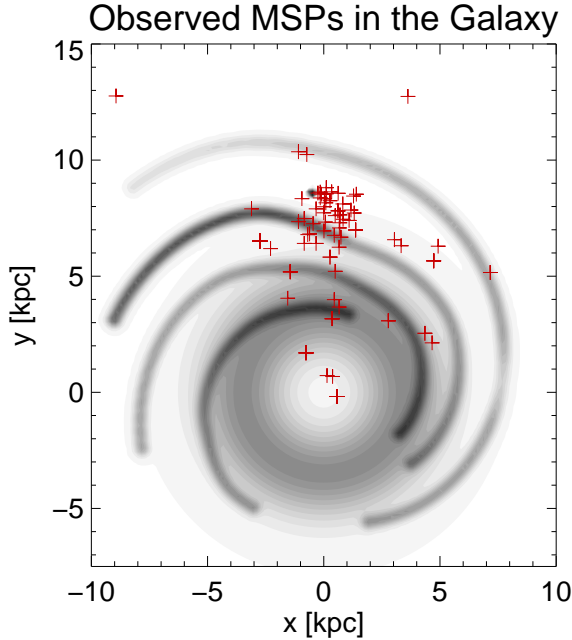,width=8.8cm}} \caption{The
distribution of the observed millisecond pulsars in the Milk Way.
The grey contour is the electron density distribution from Taylor
\& Cordes (1993). \label{image}}
\end{figure}

Recently, deep {\em Chandra} X-ray surveys of the Galactic center
(GC) revealed a multitude of point X-ray sources ranging in
luminosities from $\sim 10^{32} - 10^{35}$ ergs s$^{-1}$ (Wang,
Gotthelf, \& Lang 2002a) over a field covering a $ 2 \times 0.8$
square degree band and from $\sim 3 \times 10^{30} - 2 \times
10^{33}$ ergs s$^{-1}$ in a deeper, but smaller field of $17'
\times 17'$ (Muno et al. 2003). More than 2000 weak unidentified
X-ray sources were discovered in the Muno's field. The origin of
these weak unidentified sources is still in dispute. Some source
candidates have been proposed: cataclysmic variables, X-ray
binaries, young stars, supernova ejecta, pulsars or pulsar wind
nebulae.

EGRET on board the {\em Compton GRO} has identified a central
($<1^\circ$) $\sim 30 {\rm MeV}-10$ GeV continuum source (2EG
J1746-2852) with a luminosity of $\sim 10^{37}{\rm erg\ s^{-1}}$
(Mattox et al. 1996). Further analysis of the EGRET data obtained
the diffuse gamma ray spectrum in the Galactic center. The photon
spectrum can be well represented by a broken power law with a
break energy at $\sim 2$ GeV (see Figure 2, Mayer-Hasselwander et
al. 1998). Recently, Tsuchiya et al. (2004) have detected sub-TeV
gamma-ray emission from the GC using the CANGAROO-II Imaging
Atmospheric Cherenkov Telescope. Observations of the GC with the
air Cerenkov telescope HESS (Aharonian et al. 2004) also have
shown a significant source centered on Sgr A$^*$ above energies of
165 GeV. Some models, e.g. gamma-rays related to the massive black
hole, inverse Compton scattering, and mesonic decay resulting from
cosmic rays, are difficult to produce the hard gamma-ray spectrum
with a sharp turnover at a few GeV. However, the gamma-ray
spectrum toward the GC is similar with the gamma-ray spectrum
emitted by middle-aged pulsars (e.g. Vela and Geminga) and
millisecond pulsars (Zhang \& Cheng 2003; Wang et al. 2005). In
Figure 2, we can see that the superposed spectrum of 6000 MSPs
could significantly contribute to the observed GeV spectrum (Wang
et al. 2005).

So we will argue that there possibly exists a pulsar population in
the Galactic center region. Firstly, normal pulsars are not likely
to be a major contributor according to the following arguments.
the birth rate of normal pulsars in the Milky Way is about 1/150
yr (Arzoumanian, Chernoff, \& Cordes 2002). As the mass in the
inner 20 pc of the Galactic center is $\sim 10^8 {\rm ~M}_{\odot}$
(Launhardt, Zylka, \& Mezger 2002), the birth rate of normal
pulsars in this region is only $10^{-3}$ of that in the entire
Milky Way, or $\sim$ 1/150 000 yr. We note that the rate may be
increased to as high as $\sim 1/15000$ yr in this region if the
star formation rate in the nuclear bulge was higher than in the
Galactic field over last $10^7 - 10^8$ yr (see Pfahl et al. 2002).
Few normal pulsars are likely to remain in the Galactic center
region since only a fraction ($\sim 40\%$) of normal pulsars in
the low velocity component of the pulsar birth velocity
distribution (Arzoumanian et al. 2002) would remain within the 20
pc region of the Galactic center studied by Muno et al. (2003) on
timescales of $\sim 10^5$ yrs. Mature pulsars can remain active as
gamma-ray pulsars up to 10$^6$ yr, and have the same gamma-ray
power with millisecond pulsars (Cheng et al. 2004), but according
to the birth rate of pulsars in the GC, the number of gamma-ray
mature pulsars is not higher than 10.

On the other hand, there may exist a population of old neutron
stars with low space velocities which have not escaped the
Galactic center (Belczynski \& Taam 2004). Such neutron stars
could have been members of binary systems and been recycled to
millisecond periods, having formed from low mass X-ray binaries in
which the neutron stars accreted sufficient matter from either
white dwarf, evolved main sequence star or giant donor companions.
The current population of these millisecond pulsars may either be
single or have remained in a binary system. The binary population
synthesis in the GC (Taam 2005, private communication) shows more
than 200 MSPs are produced through recycle scenario and stay in
the Muno's region. Non-thermal emissions from MSP wind nebulae can
contribute to the X-ray sources observed by Chandra (Cheng, Taam,
Wang 2006).

\begin{figure}
\centerline{\psfig{file=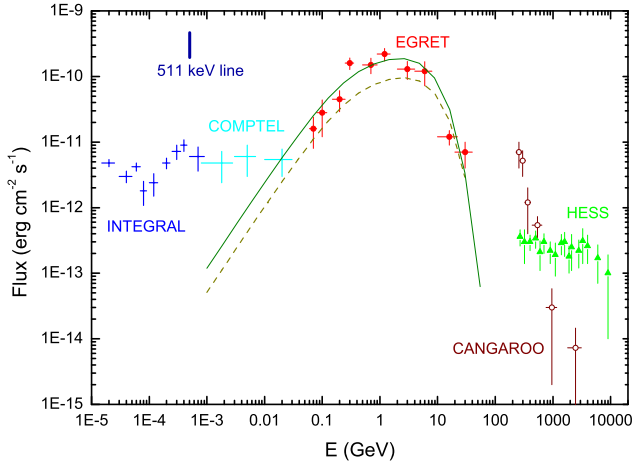,width=9.8cm }}

\caption{The diffuse gamma-ray spectrum in the Galactic center
region within 1.5$^\circ$ and the 511 keV line emission within
6$^\circ$ (from Wang 2006). The solid and dashed lines are the
simulated spectra of 6000 MSPs according to the different period
and magnetic field distributions in globular clusters and the
Galactic field respectively. \label{image}}
\end{figure}

\section{Millisecond pulsars as the positron sources}

It is well known that relativistic particles from pulsar winds
interacting with the interstellar medium form the synchrotron wind
nebulae. Chi, Cheng, \& Young (1996) proposed that the bulk of the
comic positrons could originate from pulsar winds. In this work,
we consider that the electron-positron pair production occurs in
the pulsar outer-magnetospheric region.

It has been proposed that there is a strong multipole magnetic
field near the stellar surface, although a global dipole magnetic
field gives a good description of the magnetic field far from the
star (Ruderman \& Sutherland 1975; Ruderman 1991). The typical
radius of curvature $l$ of the local magnetic field is on the
order of the crust thickness of the star (i.e. $l\sim 10^5$ cm),
which is much less than the dipole radius of curvature of dipole
field component near stellar surface. The relation between the
local multipole magnetic field and dipole field can be given by
(Zhang \& Cheng 2003) \beq B_s\simeq B_d ({R\over l})^3, \enq
where $B_d$ is the dipole magnetic field of a pulsar, $R$ is the
radius of neutron stars. For MSPs, typically $B_d \sim 10^8 -10^9$
G, $B_s\sim 10^{11}-10^{12}$ G which is much lower than the
quantum critical magnetic field $B_q\sim 4.4\times 10^{13}$ G, so
pair cascades are also efficient in the local multipole field.

The pair production mechanism is a synchrotron photon cascade in a
strong magnetic field. Photons will be converted into $e^\pm$
pairs in the local magnetic field when their energy satisfies
(Ruderman \& Sutherland 1975) \beq E\geq E_{\rm crit}\equiv
{2m_ec^2\over 15}{B_q\over B_d}({R\over l})^{-3}. \enq  The
primary $e^\pm$ from the outer-gap have the energy
$E_p=\gamma_pm_ec^2=5.7\times 10^{12}P^{1/3} {\rm eV}$, so
generally, the energies of primary curvature photons and secondary
synchrotron photons are higher than $E_{\rm crit}$, a
photon-electron cascade will start and develop until this
condition fails. At the end of a cascade, each incoming primary
electron-positron can produce, on average, \beq
N_{e^\pm}={E_p\over E_{\rm crit}}=1.9\times
10^3B_{d,9}P^{1/3}({R\over l})^3, \enq and then the total pair
production rate can be estimated as \beq
\dot{N}_{e^\pm}=f\dot{N}N_{e^\pm}=2\times
10^{33}fB_{d,9}^{10/7}P^{-8/21}({R\over l})^{30/7}{\rm s^{-1}},
\enq where $f\simeq 5.5 P^{26/21}B_{12}^{-4/7}$ is the fraction
size of the outer gap, and \beq \dot{N}=2.7\times
10^{27}P^{-2}B_{d,9}({R\over l})^3 {\rm s^{-1}} \enq is the
primary electron-positrons passing through the polar gap
(Goldreich \& Julian 1969). Taking the typical parameters $P=3$
ms, $B_d =3\times 10^8$ G, the positron injection rate for a MSP:
$\dot{N}_{e^\pm} \sim 5\times 10^{37} {\rm e^+ \ s^{-1}}$ (Wang et
al. 2006).

Since these pairs are created close to the stellar surface and the
field lines are converging, only a small fraction may keep moving
toward the star and annihilate on the stellar surface. Ho (1986)
showed that the loss cone for these pairs will approach $\pi/2$,
in other words, most pairs will be reflected by the magnetic
mirroring effect and then move toward the light cylinder. These
particles will flow out with the pulsar wind and be accelerated by
the low-frequency electro-magnetic wave.

{\it Then how many MSPs in the region of annihilation emissions?}
Figure 2 has shown that 6000 MSPs can contribute to gamma-rays
with 1.5$^\circ$, and the diffuse 511 keV emission have a size
$\sim 6^\circ$. We do not know the distribution of MSPs in the GC,
so we just scale the number of MSPs by $6000\times
(6^\circ/1.5^\circ)^2\sim 10^5$, where we assume the number
density of MSPs may be distributed as $\rho_{MSP}\propto
r_c^{-1}$, where $r_c$ is the scaling size of the GC. Then a total
positron injection rate from the millisecond pulsar population is
$\sim 5\times 10^{42}$ e$^+$ s$^{-1}$ which is consistent with the
present observational constraints. What's more, our scenario of a
millisecond pulsar population as possible positron sources in the
GC has some advantages to explain the diffuse morphology of 511
keV line emissions without the problem of the strong turbulent
diffusion which is required to diffuse all these positrons to a
few hundred pc.

\section{Discussions and conclusion}

In the present work, we suggest that there exists three possible
MSP populations: globular clusters; the Galactic field; the
Galactic Center. The population of MSPs in the GC is still an
assumption, but it seems reasonable. Importantly, the millisecond
pulsar population in the Galactic center could provide the major
sources of positrons. In addition, the scenario of a MSP
population in the GC can reasonably explain the diffuse morphology
of 511 keV line emissions.

Since there are many possible positron sources at present as noted
in \S 1. Thus, how could we distinguish the model of a millisecond
pulsar population from other models? Firstly, we can estimate the
typical spatial diffusion scale of positrons in the magnetic field
of the GC, which is given by $\lambda_{diff}\sim (r_L ct)^{1/2}$
(Wang et al. 2006), where $r_L\approx E_e/eB $ is the Larmor
radius, $E_e$ is the energy of positrons, $B\sim 10^{-5}$ G is the
average magnetic field in the GC (LaRosa et al. 2005). The average
cooling time $t$ of positrons in the GC is $\sim 10^6$ years, so
the characteristic diffusion scale is about 1 pc. Because of the
low angular resolution of SPI/INTEGRAL (about 2 degrees), we can
assume that the positrons annihilate in the local region as their
sources, i.e. the millisecond pulsars.

Therefore, we predict that the spatial intensity distribution of
the annihilation lines should follow the spatial distribution of
MSPs if a millisecond population exists in the GC. We could assume
the spatial distribution of MSPs should follow the mass
distribution of the GC though we do not know how well they follow
each other. But because the proper motion velocity of MSPs is
relatively low, we could reasonably assume that the two
distributions are quite close to each other. Then if the positron
sources originate in the MSP population, the 511 keV annihilation
line intensity would follow the mass (e.g. stars) distribution of
the Galactic center region. If the positrons originate from
supernovae or hypernovae, the 511 keV line emission could follow
the distribution of massive stars and dense molecular clouds. For
the light dark matter scenario, the annihilation emissions may
correlate to the dark matter density profile. Discrimination of
these possible correlations may be tested in the future high
resolution observations.

\vskip 0.4cm

\begin{acknowledgements}
Wang gratefully acknowledges the conference organizers and the
support by the WE-Heraeus foundation.
\end{acknowledgements}


          \clearpage


\begin{thebibliography}{}
\bibitem{aha04} Aharonian, F. et al. 2004, A\&A, 425, L13
\bibitem{acc02} Arzoumanian, Z., Chernoff, D.F. \& Cordes, J.M. 2002,
ApJ, 564, 333

\bibitem{bet02} Belczynski, K, \& Taam, R. E. 2004, ApJ, 616, 1159
\bibitem{boe04} Boehm, C. et al. 2004, Phys. Rev. Lett., 92, 101301
\bibitem{cas04} Cass\'e, M. et al. 2004, ApJ, 602, L17

\bibitem{che04} Cheng, K. S. et al. 2004, ApJ, 608, 418

\bibitem{che06} Cheng, K. S., Taam, R. E. \& Wang, W. 2006, ApJ, 641,
427

\bibitem{chd06}Cheng, K. S., Chernyshov, D. O. \& Dogiel, V. A.,
ApJ in press, astro-ph/0603659
\bibitem{ccy96} Chi, X., Cheng, K.S. \& Young, E.C.M. 1996, ApJ, 459,
L83
\bibitem{chu05} Churazov, E. et al. 2005, MNRAS, 357, 1377
\bibitem{coj69} Goldreich, P., \& Julian, W.H. 1969, ApJ, 157, 859
\bibitem{hoc86} Ho, C. 1986, MNRAS, 221, 523
\bibitem{joh73} Johnson, W.N. \& Haymes, R.C. 1973, ApJ, 184, 103
\bibitem{jea03} Jean, K. et al. 2003, A\&A, 407, L55

\bibitem{kno03} Kn\"odlseder, J. et al. 2003, A\&A, 411, L457

\bibitem{kno05} Kn\"odlseder, J. et al. 2005, A\&A, 441, 513
\bibitem{lau02} Launhardt, R., Zylka, R., \& Mezger, P. G. 2002,
A\&A, 384, 112
\bibitem{lms78} Leventhal, M., MacCallum, C.J. \& Stang, P.D. 1978,
ApJ, 225, L11
\bibitem{may98} Mayer-Hasselwander, H.A. et al. 1998, A\&A, 335, 161

\bibitem{mun03} Muno, M. P., et al. 2003, ApJ, 589, 225

\bibitem{pfa02} Pfahl, E.D., Rappaport, S., \& Podsiadlowski, P. 2002, ApJ, 571, L37

\bibitem{ran05} Ransom, S.M. et al. 2005, Science, 307, 892
\bibitem{rud75} Ruderman, M.A. \& Sutherland, P. 1975, ApJ, 196, 57

\bibitem{rud91} Ruderman, M.A. 1991, ApJ, 366, 261
\bibitem{tco93} Taylor, J.H. \& Cordes, J.M. 1993, ApJ, 411,
674
\bibitem{tee05} Teegarden, B. J. et al. 2005, ApJ, 621, 296
\bibitem{tsu04} Tsuchiya, K. et al. 2004, ApJ, 606, L115
\bibitem{wan05} Wang, W., Jiang, Z.J. \& Cheng, K.S. 2005, MNRAS, 358, 263
\bibitem{wan06} Wang, W. 2006, Chin. J. Astron. Astrophys. in press, astro-ph/0510461
\bibitem{wcj06} Wang, W., Pun, C. S.J. \& Cheng, K.S. 2006, A\&A, 446, 943
\bibitem{zhc03} Zhang, L., \& Cheng, K. S. 2003, A\&A, 398, 639
\end{thebibliography}
\end{document}